\documentclass[aps,reprint]{revtex4-1}
\usepackage{hyperref}
\usepackage{amsmath}
\usepackage{amssymb}
\usepackage[pdftex]{graphicx}
\usepackage{microtype}
\usepackage{enumerate}
\newcommand{\mrm}{\mathrm}

\begin{document}
\title{Broadband low-noise photodetector for Pound-Drever-Hall laser stabilization}
\author{Shreyas Potnis}
\author{Amar C. Vutha}
\email{vutha@physics.utoronto.ca}
\affiliation{Department of Physics, University of Toronto, 60 St.\ George Street, Toronto ON M5S 1A7, Canada}

\begin{abstract}
The Pound-Drever-Hall laser stabilization technique requires a fast, low-noise photodetector. We present a simple photodetector design that uses a transformer as an intermediary between a photodiode and cascaded low-noise radio-frequency amplifiers. Our implementation using a silicon photodiode yields a detector with 50 MHz bandwidth, gain $> 10^5$ V/A, and input current noise $< 4$ pA/$\sqrt{\mrm{Hz}}$, allowing us to obtain shot-noise-limited performance with low optical power.
\end{abstract}
\maketitle

Stabilizing lasers to high-finesse optical cavities is an ubiquitous task in atomic spectroscopy and precision optical physics. The Pound-Drever-Hall technique (PDH) is a widely used approach for laser frequency stabilization \cite{Drever1983,Black2001}. In this technique, a laser consisting of a carrier tone (at $\nu_L$) and rf phase-modulation sidebands (at $\nu_L \pm f_m$), is reflected from an optical cavity (with resonance at $\nu_c$) and detected using a fast photodetector. The output of the detector is then mixed with a reference signal at $f_m$, to obtain a baseband discriminator signal (laser frequency $\to$ voltage converter) that is proportional to $\nu_L - \nu_c$. The output of the discriminator can subsequently be used in a feedback loop to stabilize the laser to the optical cavity, or vice versa. The sensitivity of the discriminator is limited by fundamental shot noise on the detected optical power, plus technical noise added by the photodetector. Values of $f_m \gtrsim$ 5 MHz are typically used in order to obtain sufficient control bandwidth and stability in feedback loops following the discriminator. Therefore a broadband, low-noise photodetector is essential for realizing a sensitive optical discriminator with the PDH technique.

Numerous transimpedance amplifier (TIA) designs for use with photodiodes have been described in the literature (e.g., \cite{Hobbs2000,Horowitz2016},\cite{Gray1998,Schwartze2001,Bickman2005,Grote2007,Lin2012a,Eckel2012b,Zhou2014,Akiba2015}). A popular variant involves an operational amplifier configured as a current-to-voltage converter --  with appropriate techniques, low values of input current noise, along with bandwidths of $\sim$1 MHz can be obtained in such designs \cite{Hobbs2000,Horowitz2016}. However, adapting these designs to the higher frequencies used in PDH usually requires op-amps with rather large gain-bandwidth products ($\gtrsim$ 1 GHz), which can in turn lead to bandwidth limitations due to current noise ($C dV/dt$ noise) arising from the interaction between the photodiode capacitance and the input voltage noise of these op-amps (cf.\ \cite{Horowitz2016}, Section 8.11). 

\begin{figure}[h!]
\centering
\includegraphics[width=\columnwidth]{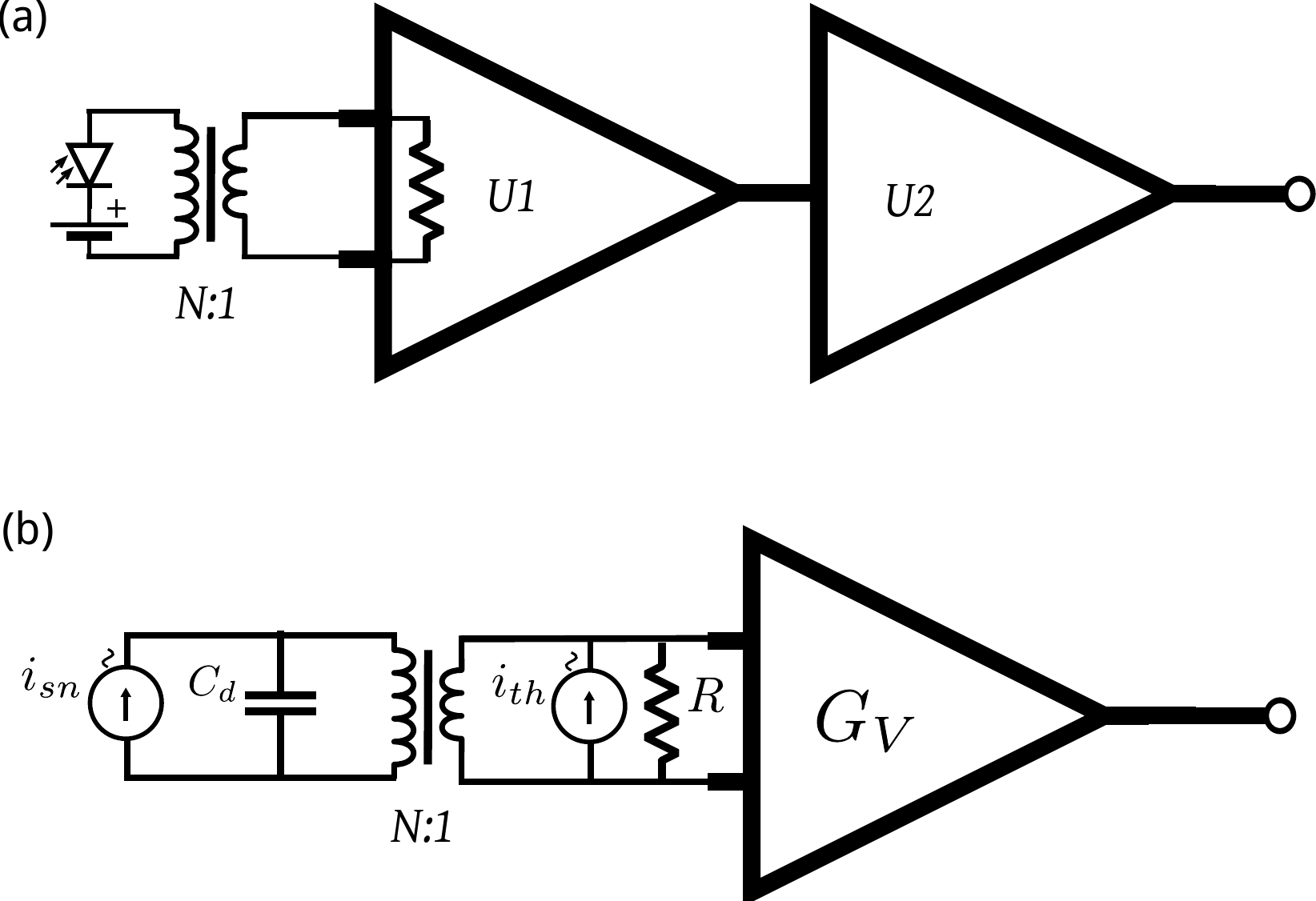}
\caption{(a) Schematic of the photodetector. U1 and U2 are low-noise rf amplifiers, with $R$ = 50 $\Omega$ input resistance and combined power gain $G_V^2 \approx 56$ dB. The transformer between the photodiode and amplifiers boosts the photocurrent shot noise above the input noise of the cascaded amplifiers. (b) Simplified noise model of the photodetector, showing the main noise sources: shot noise from the photocurrent ($i_{sn}$) and thermal noise from the rf amplifier ($i_{th}$).}
\label{fig:tia}
\end{figure}

We present a simple alternative photodetector design, shown schematically in Figure \ref{fig:tia}. A fast photodiode is reverse-biased with a battery and connected to the secondary of an $N$:1 turns ratio transformer. The primary of the transformer is directly connected to the input of two cascaded low-noise radio-frequency (rf) amplifiers (e.g., Minicircuits, ZFL-500-LN \footnote{Commercial components are identified for the sake of clarity. This does not imply that they are necessarily the best alternatives.}). 

The essential role of the transformer is to scale up the photocurrent by a factor of $N$, allowing its shot noise to dominate over the current noise from the rf amplifier. The transformer also contributes to the transimpedance gain, $G_\mrm{TI} = N R \, G_V$, where $R$ is the input resistance of the rf amplifier and $G_V$ is the voltage gain of the cascaded rf amplifiers. In addition to providing low-noise gain, the rf amplifiers are also convenient for driving coaxial cables and 50 $\Omega$ loads.

The main contributions to the input current noise density are (a) shot noise, $i_{sn} = \sqrt{2 e I_p}$, from the photocurrent $I_p$, and (b) thermal noise, $i_{th} = \sqrt{\frac{4 k_B T_n}{R}}$, arising from the input resistance ($R$ = 50 $\Omega$) of the amplifier at a noise temperature $T_n$. These contributions are schematically shown in Figure \ref{fig:tia}(b). The shot noise from the dark current through the photodiode is negligible in comparison. The total input current noise density is $i_n = \sqrt{i_{sn}^2 + \left(\frac{i_{th}}{N}\right)^2 }$. 

The bandwidth of the detector is determined by $R$, by the photodiode capacitance $C_d$, and by the transformer, which scales up the photodiode capacitance by a factor of $N^2$ as seen from the rf amplifier's input. The bandwidth expected using the model in Fig.\ 1(b) is $f_\mrm{3dB} = \frac{1}{2 \pi N^2  R C_d}$. 

An effect that can potentially limit the bandwidth is $C dV/dt$ current noise from the interaction of the photodiode capacitance with the input voltage noise of the first rf amplifier. The practical upper limit for shot-noise-limited operation is the frequency $f_c$ at which $C dV/dt$ current noise at the input, $i_c= 2 \pi f C_d N \sqrt{4 k_B T_n R}$, begins to dominate over the white noise level, $i_n$. However, it is easily seen that $f_c \geq f_\mrm{3dB}$. Therefore the $C dV/dt$ noise can always be ignored in this design, as it only makes an appreciable contribution at frequencies outside the detector's bandwidth. 

\begin{figure}[h!]
\centering
\includegraphics[width=\columnwidth]{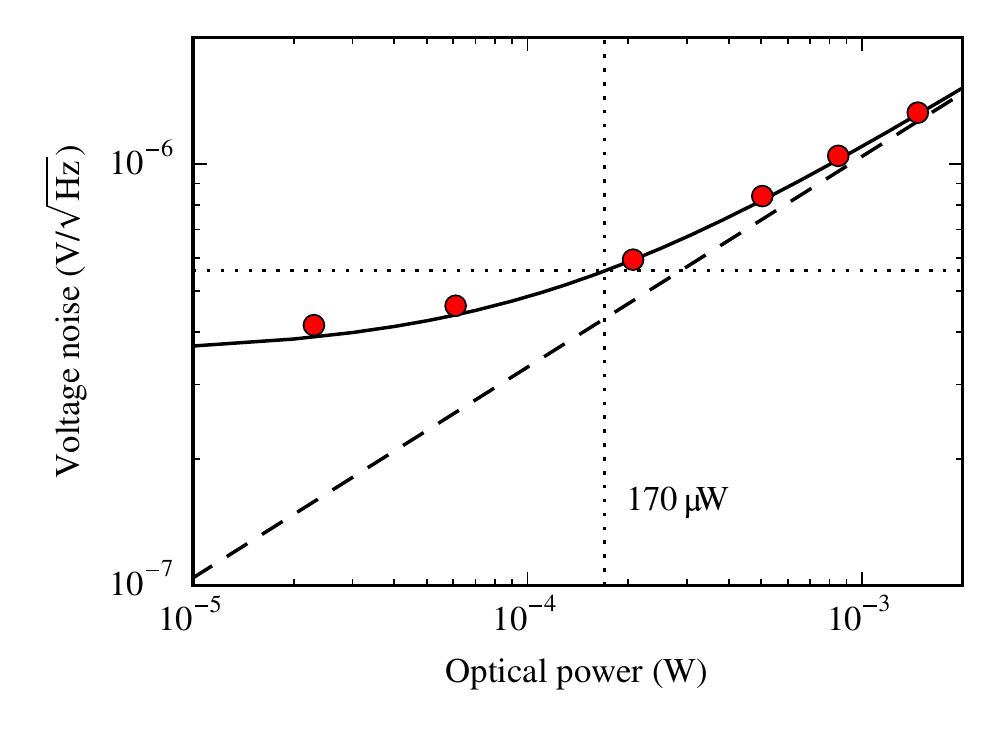}
\caption{Output voltage noise from the photodetector at $f$ = 10 MHz, measured using a spectrum analyzer. The gain is determined by fitting the measured voltage noise density, $v_n$, to the equation $v_n =\sqrt{v_{0}^{2}+2e G_\mrm{TI}^2 \mathcal{R} P}$. Here $\mathcal{R}$ is the responsivity of the photodiode, and $P$ is the optical power at 915 nm incident on the detector, as measured using a laser power meter. The fit (solid line) yields $G_\mrm{TI} = 1.3 \times 10^5$ V/A for the transimpedance gain. The dashed diagonal line shows the estimated shot noise contribution. The shot noise on the photocurrent equals the amplifier noise when 170 $\mu$W of optical power at 915 nm is incident on the detector.}
\label{fig:shot_noise}
\end{figure}

\begin{figure}[ht!]
\centering
\includegraphics[width=\columnwidth]{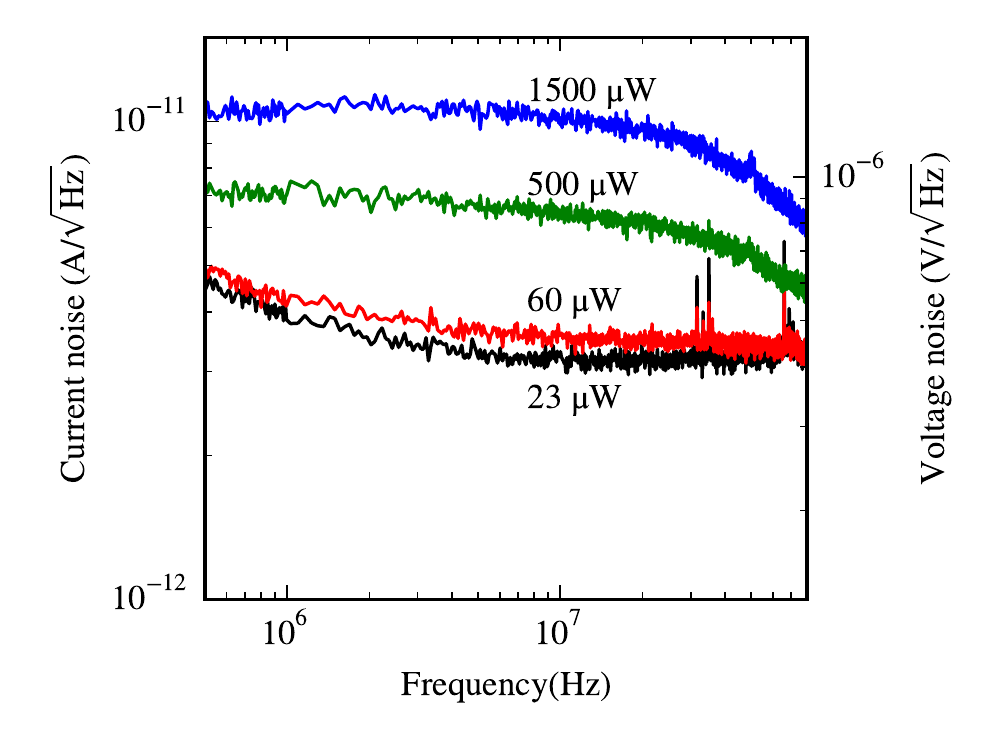}
\caption{Output voltage, and inferred input current, noise densities for some values of optical power from a 915 nm laser incident on the photodetector. The measured voltage noise is converted to input current noise using the transimpedance gain at $f$ = 10 MHz, extracted as shown in Figure \ref{fig:shot_noise}.}
\label{fig:tia_noise}
\end{figure}

\begin{figure}
\centering
\includegraphics[width=\columnwidth]{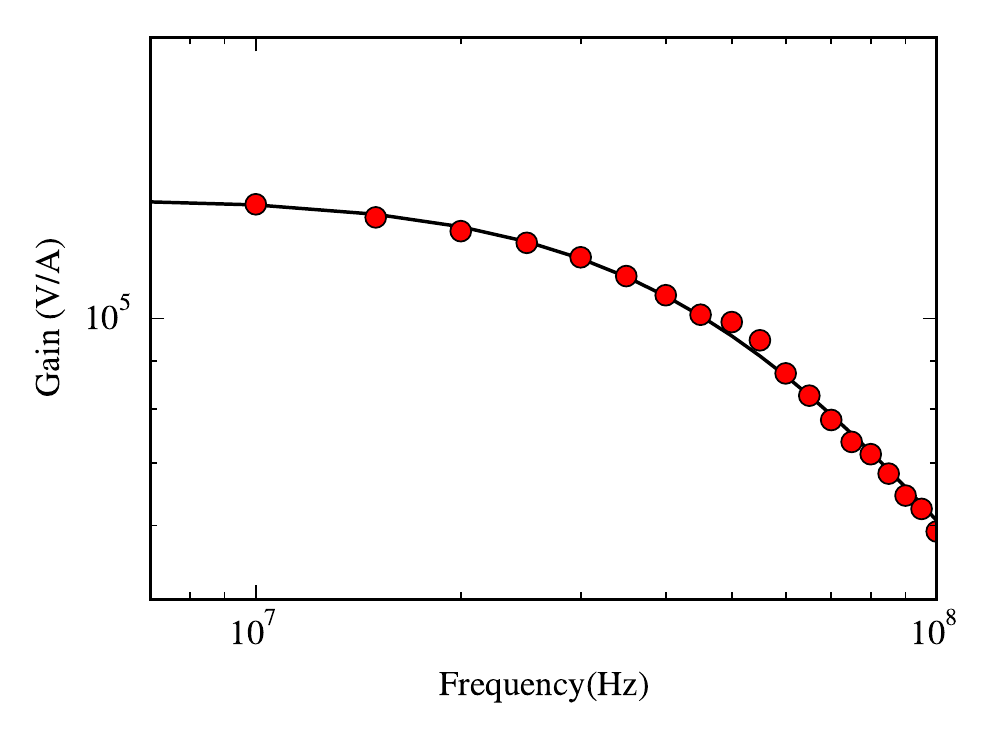}
\caption{Frequency response of the photodetector, obtained by measuring the gain (as shown in Fig.\ \ref{fig:shot_noise}) at a number of frequencies. A fit to a single-pole low-pass response (solid line) yields a bandwidth $f_\mrm{3dB} = 50 \pm 1$ MHz. The measured bandwidth agrees well with the bandwidth calculated using the model shown in Fig.\ \ref{fig:tia}.}
\label{fig:bandwidth}
\end{figure}

We implemented this design using a silicon photodiode (Thorlabs, FDS025) and a 4:1 turns ratio transformer (Minicircuits, T16-1), for use in a PDH stabilization system for a 915 nm laser. The transimpedance gain $G_\mrm{TI} = 1.3 \times 10^5$ V/A (into a 50 $\Omega$ load) was measured in two different ways: (i) from a fit to the observed voltage noise as a function of laser power as shown in Fig.\ \ref{fig:shot_noise}, using the specified responsivity of the photodiode ($\mathcal{R} = 0.22$ A/W at 915 nm), and (ii) by comparing the output of the detector to that of a commercial amplified photodiode with a specified transimpedance gain. Both methods yield consistent values for $G_\mrm{TI}$, and the measurements are in excellent agreement with the value $G_\mrm{TI,calc} = 1.26 \times 10^5$ V/A estimated from the turns ratio of the transformer and the specified gain of the rf amplifiers. 
% Thorlabs PDA10 gain = 5 kV/A, gives $G_TI = 1.58 \times 10^5$ V/A.

The measured input current noise floor, $i_n \approx 3.5$ pA/$\sqrt{\mrm{Hz}}$, shown in Fig.\ \ref{fig:tia_noise} is in reasonable agreement with $i_{n,\mrm{calc}} = 4.2$ pA/$\sqrt{\mrm{Hz}}$ estimated using the specified noise figure of the rf amplifiers and the noise model in Fig.\ \ref{fig:tia}(b). We find that the detector is shot-noise-limited for optical power levels $P \gtrsim$170 $\mu$W at 915 nm, which corresponds to photocurrents $I_p \gtrsim$ 37 $\mu$A. The detector can be shot-noise-limited for even lower optical power levels at wavelengths where the photodiode has higher responsivity.

The frequency response of the detector was measured in two different ways: first, we measured the gain using the method shown in Fig.\ 2 at a number of frequencies to obtain the gain vs.\ frequency plot shown in Fig.\ \ref{fig:bandwidth}. Next, we also measured the frequency response by varying the frequency of amplitude modulation applied to the laser (using an electro-optic amplitude modulator) and measuring the detector's output. Both methods yielded consistent results, although the latter was less precise due to fluctuations and drifts in our amplitude modulator. The measured bandwidth is $f_\mrm{3dB} =$ 50 MHz. This value is in good agreement with the value $f_\mrm{3dB,calc} \approx$ 54 MHz calculated using the specified capacitance of the photodiode ($\approx$ 4 pF with 9 V reverse bias). 
% Note the incorrect specification on Thorlabs' web page: the data that they present for FDS025 yields a capacitance of ~5 pF with 5 V bias, NOT "0.94 pF" as they incorrectly estimate using the pulse rise time (instead of using the fall time!!).

We have used these photodetectors in our laboratory to obtain shot-noise-limited PDH discriminators for stabilizing lasers to high-finesse optical cavities, typically using phase modulation frequencies $f_m \sim$ 20 MHz. Potential improvements to the design include (a) using a second-stage rf amplifier with a higher saturation level for applications involving higher optical power levels, and (b) the inclusion of a notch filter between the first- and second-stage amplifiers to suppress harmonics of the modulation frequency. The ease of implementation of this fast photodetector design, and the ability to achieve shot-noise-limited performance with low optical power, could be useful in a variety of experiments that use the Pound-Drever-Hall technique.

We acknowledge helpful suggestions from the reviewers of the manuscript. This work is supported by the University of Toronto, and by a Branco Weiss Fellowship from Society in Science, ETH Zuerich.

\bibliography{xformer_tia}
\end{document}